%% file: main.tex
\documentclass[lettersize,conference]{IEEEtran}
\usepackage{amsmath,amsfonts}
\usepackage{algorithmic}
\usepackage{algorithm}
\usepackage{array}
\usepackage[caption=false,font=normalsize,labelfont=sf,textfont=sf]{subfig}
\usepackage{textcomp}
\usepackage{stfloats}
\usepackage{xurl}
\usepackage{verbatim}
\usepackage{graphicx}
\usepackage{cite}
\usepackage[hidelinks]{hyperref}
\usepackage{color}
\usepackage{multicol}

\begin{document}

\title{PaceMaker: A Practical Tool for Pacing Video Games}

\author{
\IEEEauthorblockN{Julian Geheeb}
\IEEEauthorblockA{
\textit{Technical University of Munich}\\
Munich, Germany \\
julian.geheeb@tum.de}
\and
\IEEEauthorblockN{Daniel Dyrda}
\IEEEauthorblockA{
\textit{Technical University of Munich}\\
Munich, Germany \\
daniel.dyrda@tum.de}
\and
\IEEEauthorblockN{Sebastian Geheeb}
\IEEEauthorblockA{
\textit{Technical University of Munich}\\
Munich, Germany \\
sebastian.geheeb@tum.de}
}

\markboth{Journal of \LaTeX\ Class Files,~Vol.~14, No.~8, August~2021}%
{Shell \MakeLowercase{\textit{et al.}}: A Sample Article Using IEEEtran.cls for IEEE Journals}


\maketitle
\IEEEoverridecommandlockouts
\IEEEpubid{\makebox[\columnwidth]{ 979-8-3503-5067-8/24/\$31.00~\copyright2024 IEEE \hfill} 
\hspace{\columnsep}\makebox[\columnwidth]{ }}
\IEEEpubidadjcol

\input{sections/abstract.tex}

\begin{IEEEkeywords}
Pacing, Toolkit, Pacing Diagrams, Game Design, Game Engineering, Play, Player Experience.
\end{IEEEkeywords}

\section{Introduction}\label{sec:introduction}
Pacing is a concept that can be encountered in various fields.
On the one hand, we have media like literature, film, or television, where pacing considerations influence how fast and intense given parts of a story should feel for the consumer.
On the other hand, we have sports, where pacing deals with the distribution of intensity of the athlete's activity to avoid running out of energy before reaching the goal.
Video games typically combine the passively consumed storytelling aspects of the former with the active elements of the latter to create an interactive experience.
This presents a number of unique challenges when it comes to pacing considerations during the game development process, as both aspects need to be coordinated simultaneously.
Furthermore, a non-linear potential is created by the possibility space of games~\cite{bogost2008rhetoric}, and games are often iterated on during development and playtesting.

All of these characteristics make it hard to adopt workflows from other fields that are used to create the desired pacing.
Conventional planning on paper can quickly become intricate with an exponentially growing number of paths to consider.
Furthermore, revisiting and refining previous notes and designs during iterations can be challenging and time-consuming.
Therefore, this research aims to build a toolkit that enables common workflows while dealing with the challenges mentioned above.
The toolkit, \textit{PaceMaker}, allows the user to design a non-linear experience chart and subsequently plot relevant information like intensity or gameplay category of each node along a path on the chart.
The current prototype provides a proof of concept that shows the potential of such a tool.

The contributions of this research are as follows:
\begin{itemize}
    \item In \autoref{sec:pacing}, we present the results of our initial research into pacing and related workflows.
    \item Afterward, we provide a description of the current implementation of PaceMaker in \autoref{sec:pacemaker}.
    \item Finally, we evaluate the toolkit by providing results in \autoref{sec:results}, discussing the findings in \autoref{sec:discussion}, comparing it to previous studies in \autoref{sec:related_work}, and expanding upon the design in \autoref{sec:future_work}.
\end{itemize}

\section{Pacing}\label{sec:pacing}
Before implementing PaceMaker, we investigated pacing in video games and related workflows.
We searched for sources within the academic field and on the platform \textit{Gamedeveloper}. 
We believe that there is much to be learned from both the academic and professional domain~\cite{greenwood2021understanding}.
During this research, we derived a definition of pacing and identified pacing diagrams as the most common artifact that informs the design process of pacing.
We subsequently derived our toolkit requirements from our findings and from the problem statement itself.

\subsection{Definition}
This section presents our definition of pacing, which was derived by identifying similarities among the pacing definitions of \cite{gyllenback2020pacing1,hill2013ticktickboom,ho2017pacing,lopez2009pacing1,costiuc2019uncharted1,bagus2022pacing,hoffstetter2013combatpacing,linehan2014learning,venturelli2009space,davies2009gamepace}.
Pacing describes the rhythm that results from the recurring patterns of rhythmic parameters in time.
Rhythmic parameters are divided into \textit{artifact} and \textit{experience parameters}.
Artifact parameters describe tangible values and properties of the game itself, which can be influenced by the game designer, e.g. the number of enemies per level.
Experience parameters describe an experience of the player while interacting with the game, e.g. narrative intensity, and are influenced by the game itself through artifact parameters.
We found that most rhythmic parameters mentioned fall into one of two categories: numerical or categorical.
Both play time and event time~\cite{nitsche2007mapping} are considered, but event time seems to be the focus of most authors.
When talking about event time, many authors use the term \textit{beat} to describe the basic unit of time, where a beat can be on any level of detail.

Pacing can be divided into different components that make up a game.
The discussion often centers around narrative \cite{costiuc2019uncharted1,gyllenback2020pacing1,hill2013ticktickboom,ho2017pacing,lopez2009pacing1} and gameplay pacing \cite{bagus2022pacing,hoffstetter2013combatpacing,linehan2014learning,venturelli2009space,davies2009gamepace}, but pacing can also be influenced by music, sound design, visuals, and more \cite{costiuc2019uncharted1}.
A mindmap illustrating our definition can be seen in \autoref{fig:pacing_mindmap}.

\subsection{Pacing Diagrams}
A common approach for designers is to use a visual graphic to describe the pacing of a game (see \cite{coulianos2011pacing, bagus2022pacing, hoffstetter2013combatpacing, linehan2014learning, costiuc2019uncharted1,lopez2009pacing1}), which we call a pacing diagram.
A pacing diagram is a tool used by designers to plan the pacing of a linear sequence of play activities.
The diagram plots a numerical or type parameter against event time on the x-axis and the rhythmic parameter value on the y-axis.
The primary experience parameter is intensity, although others such as engagement, threat, movement impetus, and tempo may also be considered.
When visualizing artifact parameters, the most frequently used one is the gameplay type, although others such as the required number of actions are also possible. 
Pacing diagrams can be used for small-scale segments, such as levels, and larger segments, such as the entire game. 
The pacing diagram provides a formal description of a linear play experience.

\input{sections/pacemaker.tex}

\input{sections/results.tex}

\input{sections/discussion.tex}

\section{Related Work}\label{sec:related_work}

Bagus et al.~\cite{bagus2022pacing} describe a dungeon level generator that considers pacing patterns intended by the game designer.
The toolkit's input is an original level of which the designer can see and set pacing patterns and constraints of all rooms.
Afterwards, the toolkit creates a population of rooms, and the designer can choose the preferred one.
Just like PaceMaker, the generator makes use of line diagrams to visualize pacing parameters across rooms for the designer. 
The toolkit focuses on the three parameters \textit{threat}, \textit{impetus}, and \textit{tempo}, while our focus lies on intensity.
Similarly to our division of rhythmic parameters, there are parameters, like threat, that are calculated through formulas of other parameters, like \textit{damage taken}, \textit{movement}, and \textit{distance}.
Our research differs in two important aspects.
First of all, PaceMaker is purely a design and planning tool.
Second, our toolkit is universal in terms of gameplay, which means it is not restricted to dungeon level generation only.

\textit{Launchpad}~\cite{smith2010launchpad} is a level generator for 2D platforming games that uses a rhythm-based approach.
A level is formalized through rhythm groups, which are "short, nonoverlapping sets of components that encapsulate an area of challenge"~\cite[p.5]{smith2010launchpad}, which is close to what we call beats.
The generation is controlled by a set of parameters that influence the level pacing and geometry, similar to our artifact parameters.
After a level has been generated, its \textit{leniency} is evaluated, which can be considered as inverse intensity.
Our toolkit separates itself from \textit{Launchpad} in the fact that PaceMaker is a planning tool for all sorts of games instead of a content generation tool for a specific genre.
The creators of \textit{Launchpad} later developed a mixed-initiative toolkit called \textit{Tanagra}~\cite{smith2011tanagra}, which generates levels similarly, but also allows a human designer to finetune them.
While this can be considered a planning tool with content generation on top, PaceMaker differs in its flexibility when it comes to the game genre.

A system used in a successful commercial game, \textit{Left 4 Dead}~\cite{left4dead}, is its \textit{AI Director}~\cite{booth2009left4dead}.
One of the \textit{AI Director}'s goals is to generate dramatic game pacing.
This is accomplished by an adaptive dramatic pacing algorithm, which creates peaks and valleys of intensity.
The intensity is estimated per player through parameters like \textit{damage taken} or \textit{incapacitated player state}, which are artifact parameters in our tool.
It then tracks the maximum intensity of all players and removes major threats if it is too high or creates an interesting population of threats otherwise.
The intensity is also plotted along with the population density.
This formalized approach to pacing and intensity is very similar to ours.
However, unlike with PaceMaker, the course of the pacing curve is not explicitly designed.
Rather, \textit{Left 4 Dead}'s pacing is created dynamically based on predefined rules consisting of a four-step repeating cycle.

\section{Conclusion \& Future Work}\label{sec:future_work}
In this research, we researched pacing and related design processes in video games.
We subsequently created a toolkit, PaceMaker, that offers functionality for the most relevant findings, like support for rhythmic parameters and pacing diagrams.
Additionally, we considered unique challenges occurring during game development and implemented an experience chart to deal with the non-linear potential of games and experience specifications for iteration and flexibility.
While the results give rise to some concerns about the efficiency of the toolkit, its expressiveness, and the general feedback from the interviews are a promising outcome, supporting the need for such a tool.

Future iterations will show whether the efficiency and usability can be increased sufficiently and if it can be adopted in a game development environment.
For this, we aim to investigate the issues discussed in \autoref{sec:discussion} by implementing additional statechart features and improving the usability and user experience of PaceMaker. 
Subsequently, we plan to conduct a comprehensive user study to arrive at more definitive conclusions.


 

\bibliographystyle{IEEEtran}
\bibliography{bibliography}

\vfill

\end{document}

%% file: sections/abstract.tex
\begin{abstract}
Designing pacing for video games presents a unique set of challenges.
Due to their interactivity, non-linearity, and narrative nature, many aspects must be coordinated and considered simultaneously.
In addition, games are often developed in an iterative workflow, making revisions to previous designs difficult and time-consuming.
In this paper, we present PaceMaker, a toolkit designed to enable common design workflows for pacing while addressing the challenges above.
We conducted initial research on pacing and then implemented our findings in a platform-independent application that allows the user to define simple state diagrams to deal with the possibility space of games.
The user can select paths on the directed graph to visualize a node's data in diagrams dedicated to intensity and gameplay category.
After implementation, we created a demonstration of the tool and conducted qualitative interviews.
While the interviews raised some concerns about the efficiency of PaceMaker, the results demonstrate the expressiveness of the toolkit and support the need for such a tool.
\end{abstract}

%% file: sections/pacemaker.tex
\section{PaceMaker}\label{sec:pacemaker}
PaceMaker was built as a web application to be platform- and engine-independent.
Web apps can be deployed and accessed through browsers on any machine, as well as built as cross-platform desktop applications with tools such as \textit{Electron}.
While many alternative frameworks in the web development domain are reasonable choices, we decided to go with \textit{Vue3}, as provides an approachable, versatile, and lightweight option.
For the first iteration of PaceMaker, we decided to focus on the functional requirements.
While usability and other non-functional requirements are important aspects of a tool's success, first prototypes are usually meant to be a functional proof of concept.
Based on the findings presented in \autoref{sec:pacing}, it is clear that the requirements include the ability to handle rhythmic parameters and present them visually through pacing diagrams for a linear experience.
Since intensity and gameplay type are the most common parameters in each category, we focused on implementing them first.

\subsection{Experience Chart}
The experience chart models the non-linear experience of the player in the form of simple Statecharts~\cite{harel1987statecharts}.
By using Statecharts, game developers can formally approach the design and interaction complexity of games, mapping out the expansive possibility space that defines the player's experience.
A node in the experience chart is called a \textit{beat} and represents a part of the game that results in an experience for the player, like a sequence of actions, a section, or a whole level. 
An edge represents the dependencies between the different parts of the game, where a directed edge from \textit{Beat 1} to \textit{Beat 2} means that \textit{Beat 2} follows after \textit{Beat 1}.
The current implementation only allows for atomic states without nesting or concurrency.

\begin{figure}[!t]
    \centering
    \includegraphics[width=\columnwidth]{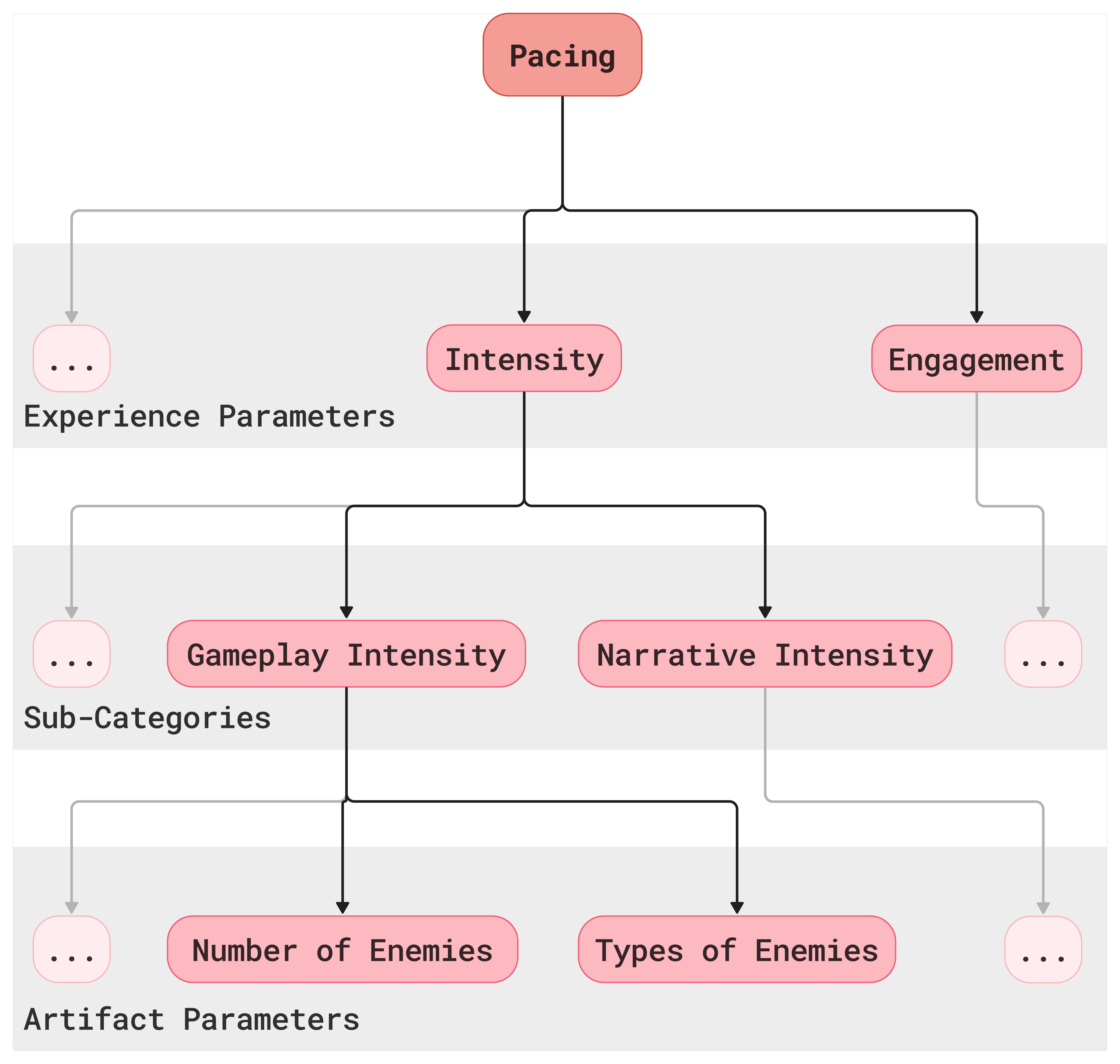}
    \caption{A pacing mindmap of the final definition. Here, the sub-categories are applied between the artifact and experience parameters. However, this sub-categorization can be applied on any level.}
    \label{fig:pacing_mindmap}
\end{figure}

\subsection{Experience Specification}
An experience specification can be assigned to a beat, which contains a list of properties that define the player experience as follows: 
\begin{itemize}
    \item[] \textbf{Name}: The name of the specification
    \item[] \textbf{Description}: The description of the part of the game that results in the experience
    \item[] \textbf{Narrative Intensity}: A numerical parameter that can assume any number value in $[0, 100]$
    \item[] \textbf{Gameplay Intensity}: A numerical parameter that can assume any number value in $[0, 100]$
    \item[] \textbf{Intensity}: A numerical parameter representing the overall intensity, computed by the average of narrative and gameplay intensity
    \item[] \textbf{Gameplay Category}: A categorical parameter that can assume any string value defined by the user
    \item[] \textbf{Expected Playtime}: A mapping between event time and play time
\end{itemize}
The properties \textit{Narrative Intensity}, \textit{Gameplay Intensity}, Intensity, and \textit{Gameplay Category} were chosen based on \autoref{sec:pacing}.
While this is not an extensive list of properties that describe an experience, it covers the most common ones.
As discussed earlier, it is difficult to quantify experience parameters like intensity. 
However, we can use artifact parameters to approximate experience parameters, since the experience is directly influenced by the game artifact itself.
Therefore, the vision of this part of the system is that the user can define artifact parameters based on their needs, and use them to compute a value to approximate the experience.
To demonstrate this relation, we divided the \textit{Intensity} property of our experience specification into \textit{Narrative} and \textit{Gameplay Intensity}, which, according to \autoref{sec:pacing}, coincide with the most relevant components of a game.
Expected playtime was added to later visualize differences between play time and event time~\cite{nitsche2007mapping}.

\subsection{Path Selection}
The path selection allows the user to select a path on the experience chart along which the data is to be displayed later (\autoref{sec:pacing_diagrams}).
The user can select a start beat, intermediate beats, and an end beat.
The Dijkstra algorithm is used to calculate a path along those beats, which gives the user additional information, such as showing the critical path, or whether the selected beats can even form a path in that order.
If a path exists, it is highlighted for the user.
Furthermore, snapshots of paths can be taken to select and compare multiple paths simultaneously.

\subsection{Pacing Diagrams}\label{sec:pacing_diagrams}
Pacing diagrams visualize the data of experience specifications along a selected path of beats of the experience chart.
The diagrams allow for basic functionality like hiding datasets, zooming to change the level of detail, and downloading the data as .svg, .png, or .csv files for further processing.

\subsubsection{Intensity Diagrams}
Intensity diagrams are divided into two categories.
In the first category, an intensity data point is added for each beat with an assigned experience specification, representing event time.
In the second category, intensity data points are added based on a beat's expected playtime, representing playtime.
Furthermore, the user can switch the intensity setting between \textit{Computed}, \textit{Gameplay}, \textit{Narrative}, and \textit{All} to display the corresponding properties.

\subsubsection{Gameplay Category Diagrams}
Gameplay category diagrams display the gameplay categories along the selected paths on a timeline.
Additionally, the user can switch the timeline settings between \textit{Beat}, where each beat has the same length, and \textit{Time}, where the length of the beat is determined by the expected playtime of the assigned experience specification.

%% file: sections/results.tex
\section{Results}\label{sec:results}
To evaluate the toolkit, we created meaningful artifacts and conducted an initial user study in the form of expert interviews.
The following sections provide a detailed description of the steps taken to create these results, as well as the results themselves.

\subsection{Walkthrough Demonstration}
\begin{figure}[!t]
    \centering
    \includegraphics[width=\columnwidth]{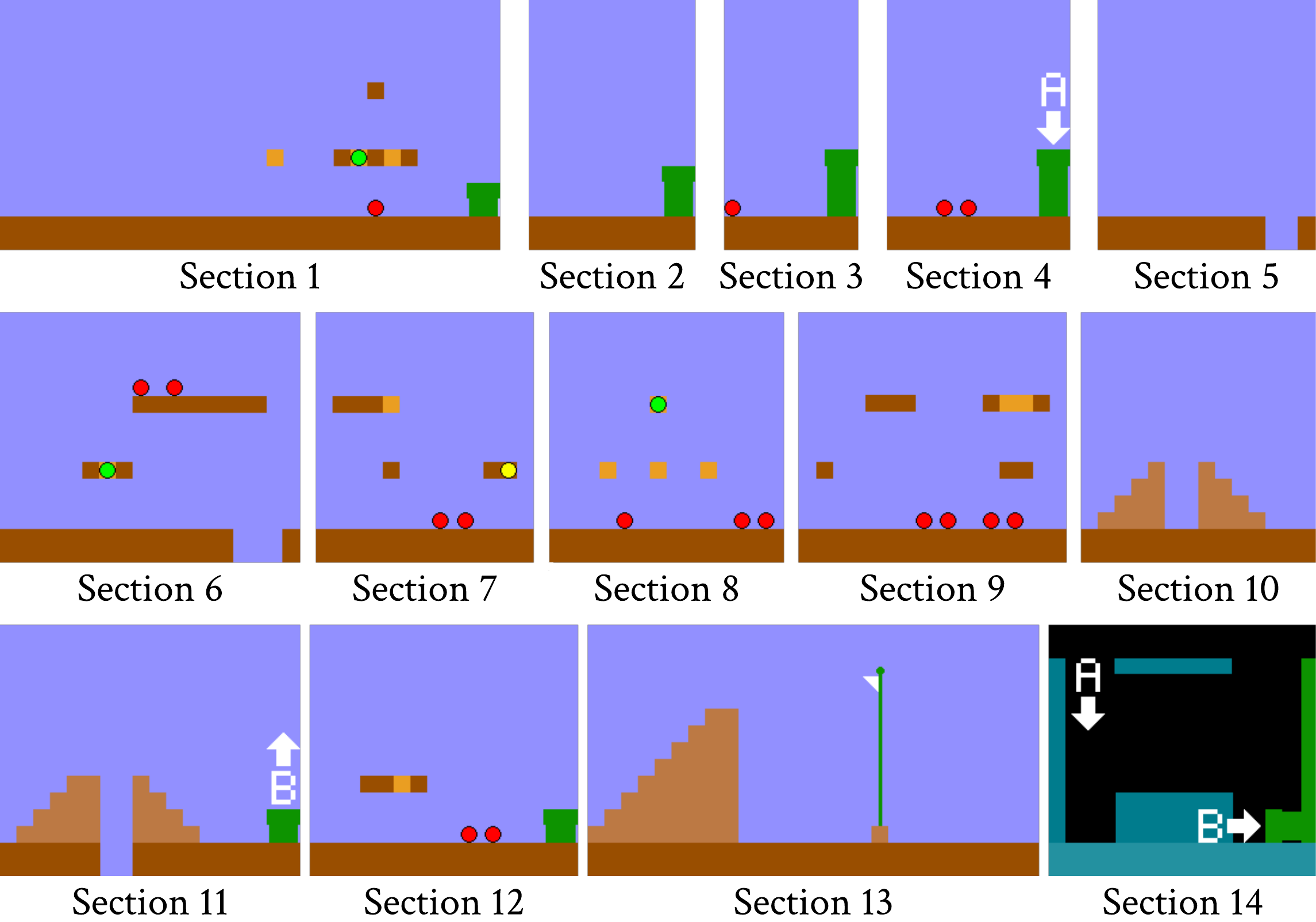}
    \caption{A replica of \textit{World 1-1} of \textit{Super Mario Bros.} divided into 14 sections. Red dots correspond to enemies, green dots correspond to common power-ups, and the yellow dot corresponds to the star power-up.}
    \label{fig:world_1_1}
\end{figure}

A demonstration of how the toolkit can be used explores the \textit{design space} and investigates the toolkit's \textit{ceiling}~\cite{ledo2018evaluation}.
Although demonstrations might not fully showcase the height of the ceiling, they serve as an indication of the toolkit's potential complexity and the range of possible solutions~\cite{ledo2018evaluation}.
For this walkthrough, we create a simple, non-extensive model of \textit{World 1-1}'s game space and pacing of \textit{Super Mario Bros.}~\cite{supermariobros} with sufficient complexity.
For this, we divide the level into 14 sections (\autoref{fig:world_1_1}).
The responsibility of creating such a model usually lies with the designer who utilizes the toolkit and can be arbitrarily simple or complex.

\subsubsection{Modeling Experience Specifications}
Initially, the user should be able to assign an intensity and gameplay type value to each experience specification.
This can be accomplished by any means the user deems appropriate, such as determining the value based on personal perception or analyzing feedback from playtests during subsequent iterations.
In this instance, we utilize a model based on artifact parameters, which are defined as follows:
\begin{itemize}
    \item[] $Tiles$: Number of tiles
    \item[] $Enemies$: Number of enemies
    \item[] $State_e$: Enemies' state: ground: 1, platform: 0.5
    \item[] $State_p$: Player's state: small: 1, big: 0.5, flower: 0.25, star: 0
    \item[] $Jump_v$: The number of vertical tiles required to jump
    \item[] $Jump_h$: The number of horizontal tiles required to jump
\end{itemize}
The intensity value $f_i$ is then calculated as follows:
\[f_j = Jump_v + Jump_h \cdot 3\]
\[f_e = State_e \cdot \frac{Enemies}{Tiles} \cdot 400 \cdot State_m\]
\[f_i = round(f_j + f_e)\]
Similarly, the gameplay type of each experience specification can be decided by some simple rules in this priority:
\begin{itemize}
    \item[] \textbf{Reward}: All sections where the player is in star state, Section 13 (Goal), Section 14 (Secret Area) 
    \item[] \textbf{Pit}: All sections where there is a pit the player can fall into and die
    \item[] \textbf{Enemy}: All sections where there is at least one enemy
    \item[] \textbf{Platforming}: All remaining sections
\end{itemize}
In most cases, we set both intensity values of each experience specification to $f_i$ in relation to the artifact parameters of the corresponding experience.
For specifications with the type \textit{Reward}, we set \textit{Narrative Intensity} to 100.
Finally, we set \textit{Expected Playtime} to the number of tiles in seconds.

\subsubsection{Creating the Experience Chart}
\begin{figure*}[!t]
    \centering
    \includegraphics[width=0.9\textwidth]{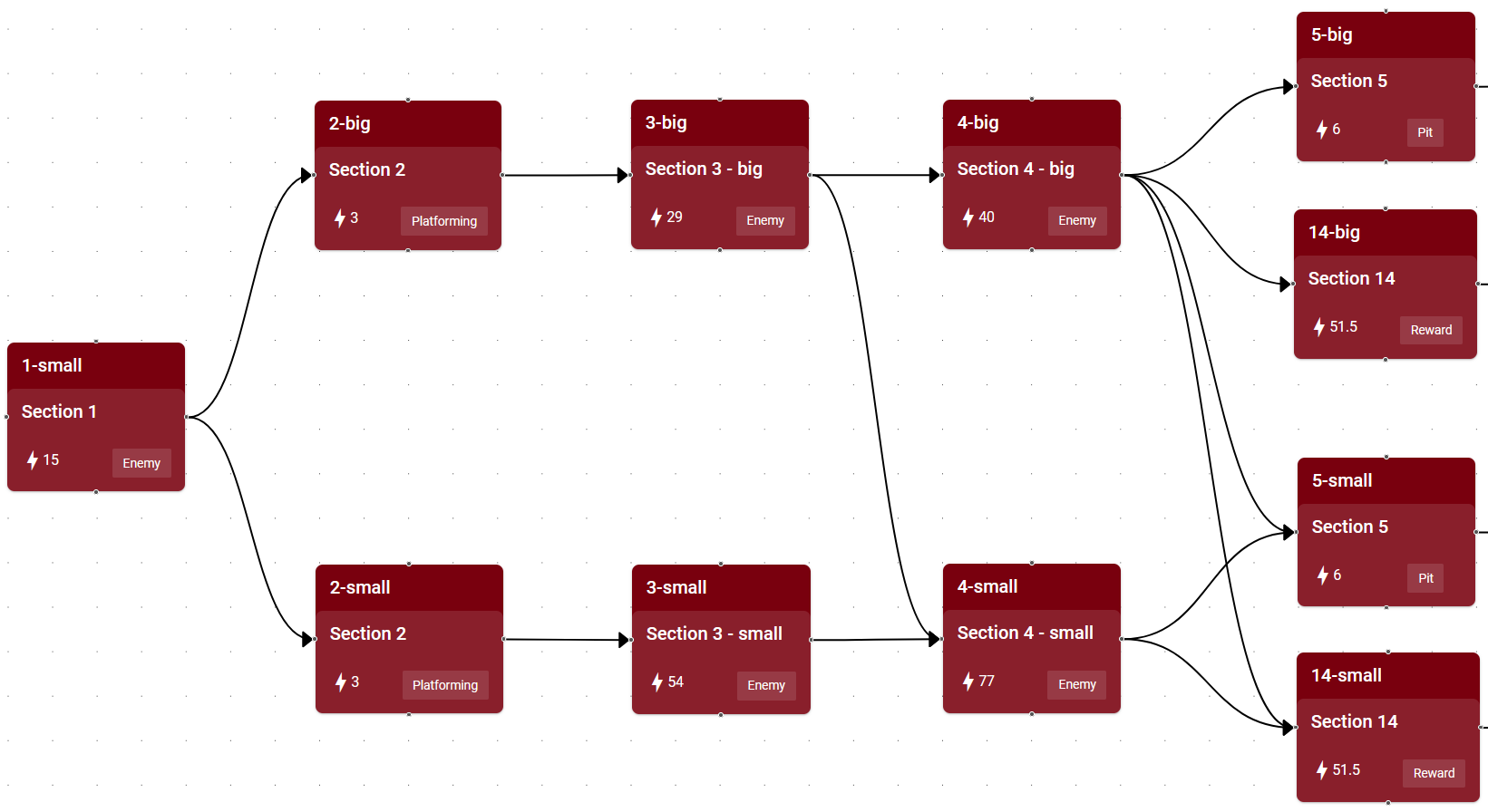}
    \caption{The first five sections modeled in PaceMaker's experience chart during our walkthrough demonstration. A beat displays the name of the beat and a preview of the experience specification in the form of its name, the intensity value, and the gameplay category.}
    \label{fig:mario_5_sections}
\end{figure*}

Now the user can start creating the experience chart.
For the first part of this description, the reader can reference \autoref{fig:mario_5_sections}.
Assuming the player plays the game for the first time, the initial player state is 'small'.
For simplicity, we only consider the player state while entering a section.
Furthermore, we identify beats by the combination of their section and the player state, e.g. '1-small' stands for the beat of Section 1 with the 'small' player state.
We create a beat for Section 1 and assign a specification based on our rules.
In Section 1, the player can find a power-up.
The player can therefore enter Section 2 in two different states, so we create two separate beats, '2-big' and '2-small', and add the edges '1-small $\rightarrow$ 2-small' and '1-small $\rightarrow$ 2-big'.
Since the player state does not influence the gameplay of Section 2, we only need to create one experience specification and can assign it to both beats.
In Section 2, it is not possible to change the player state.
Consequently, we create one following beat for each beat of section two and assign an experience specification accordingly.
In Section 3, the player can lose the player state 'big', so we create two beats for Section 4, and add the edges '3-small $\rightarrow$ 4-small', '3-big $\rightarrow$ 4-small', and '3-big $\rightarrow$ 4-big'.
In Section 4, the player can enter the pipe to reach Section 14 or continue the level normally to Section 5, which results in four new beats.
Furthermore, the player can lose the player state 'big', which leads to '4-big' being connected to all new beats, while '4-small' only has edges to '5-small' and '14-small'.

\begin{figure*}
    \centering
    \includegraphics[width=\textwidth]{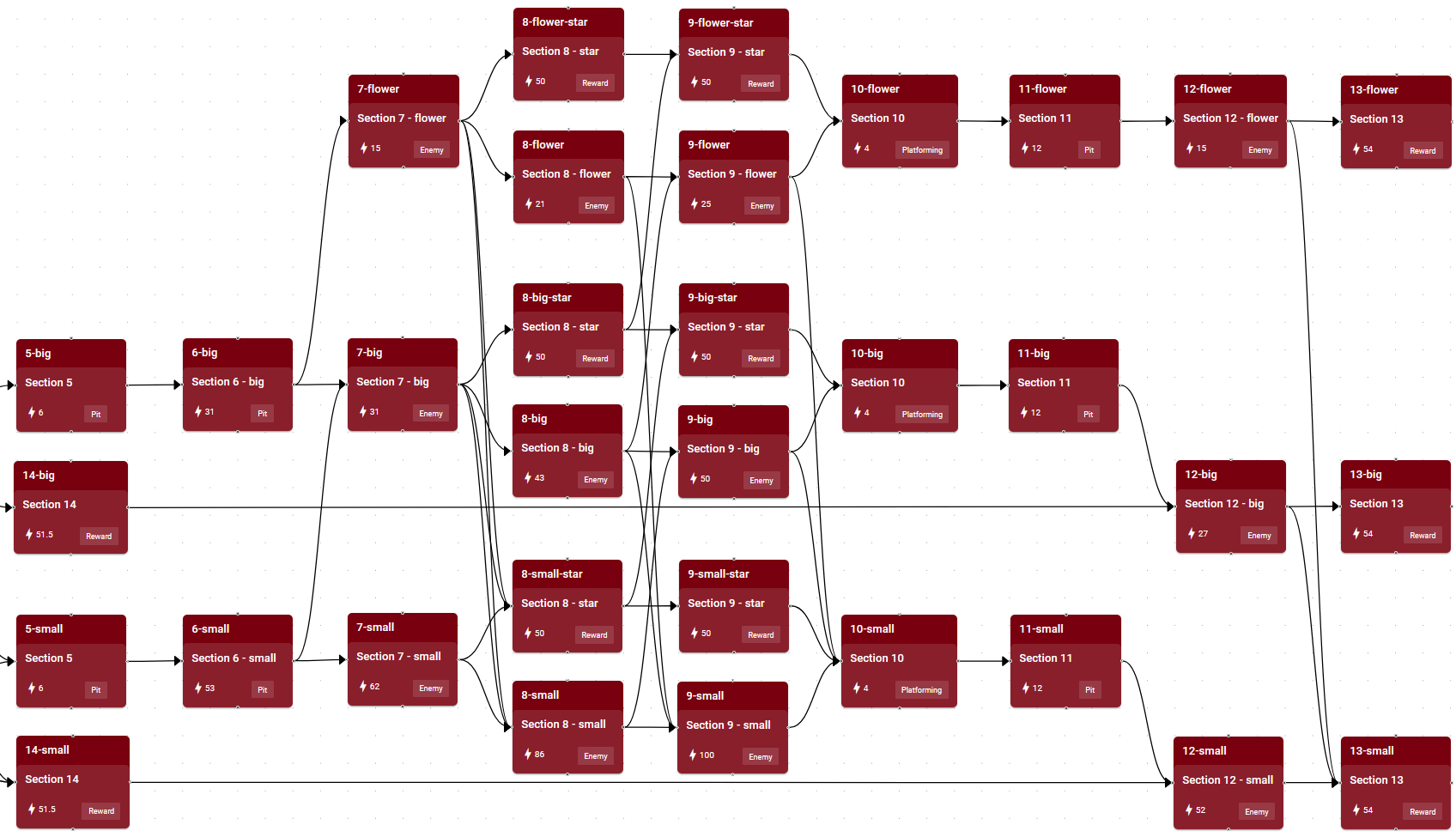}
    \caption{The remaining section to complete the model in PaceMaker's experience chart during our walkthrough demonstration. The leftmost beats overlap with \autoref{fig:mario_5_sections}'s rightmost beats for continuity.}
    \label{fig:mario_9_sections}
\end{figure*}

The remaining part of the experience chart (\autoref{fig:mario_9_sections}) was generated using the same logic, except for the star state, which is only active for a specific duration during the game.
In our approach, we assume that if the player activates the 'star' power-up, it remains active for Section 8 and Section 9, after which the player returns to their original state before the power-up.

\subsubsection{Visualizing Experiences Through Pacing Diagrams}
\begin{figure*}
    \includegraphics[trim=75 517 928 10, clip, width=0.33\textwidth]{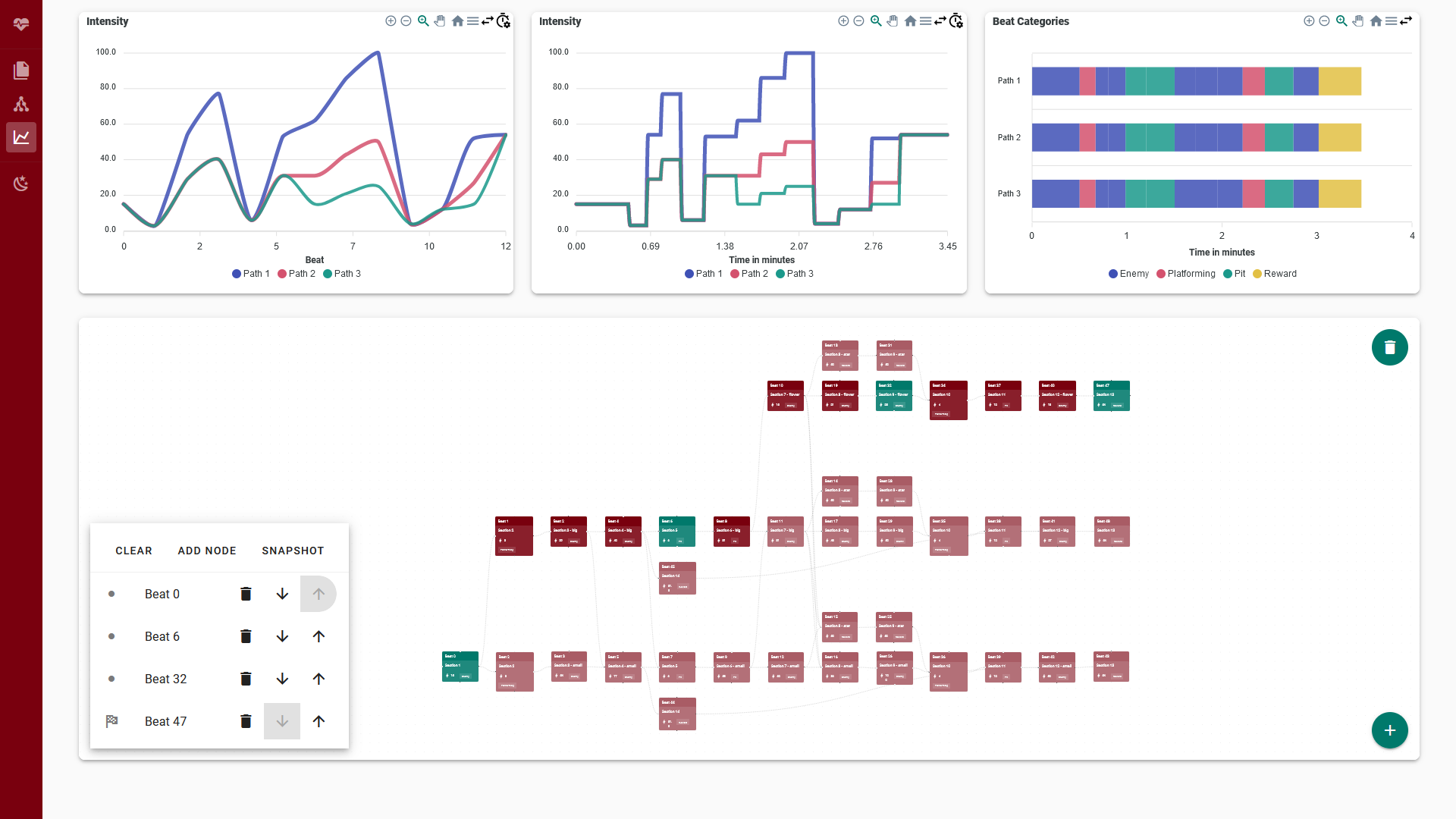}
    \includegraphics[trim=75 517 928 10, clip, width=0.33\textwidth]{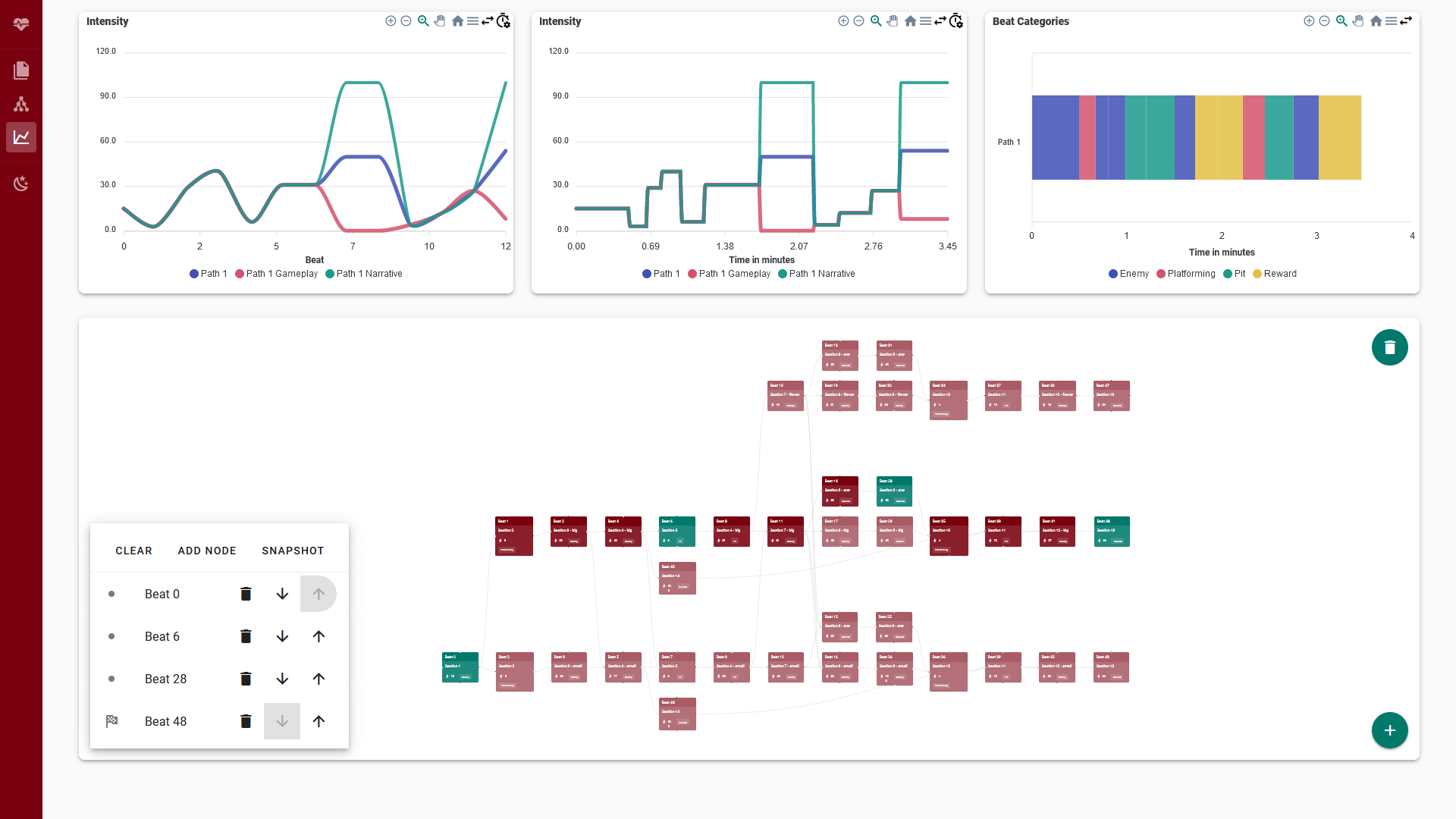}\vspace{1cm}
    \includegraphics[trim=75 517 928 10, clip, width=0.33\textwidth]{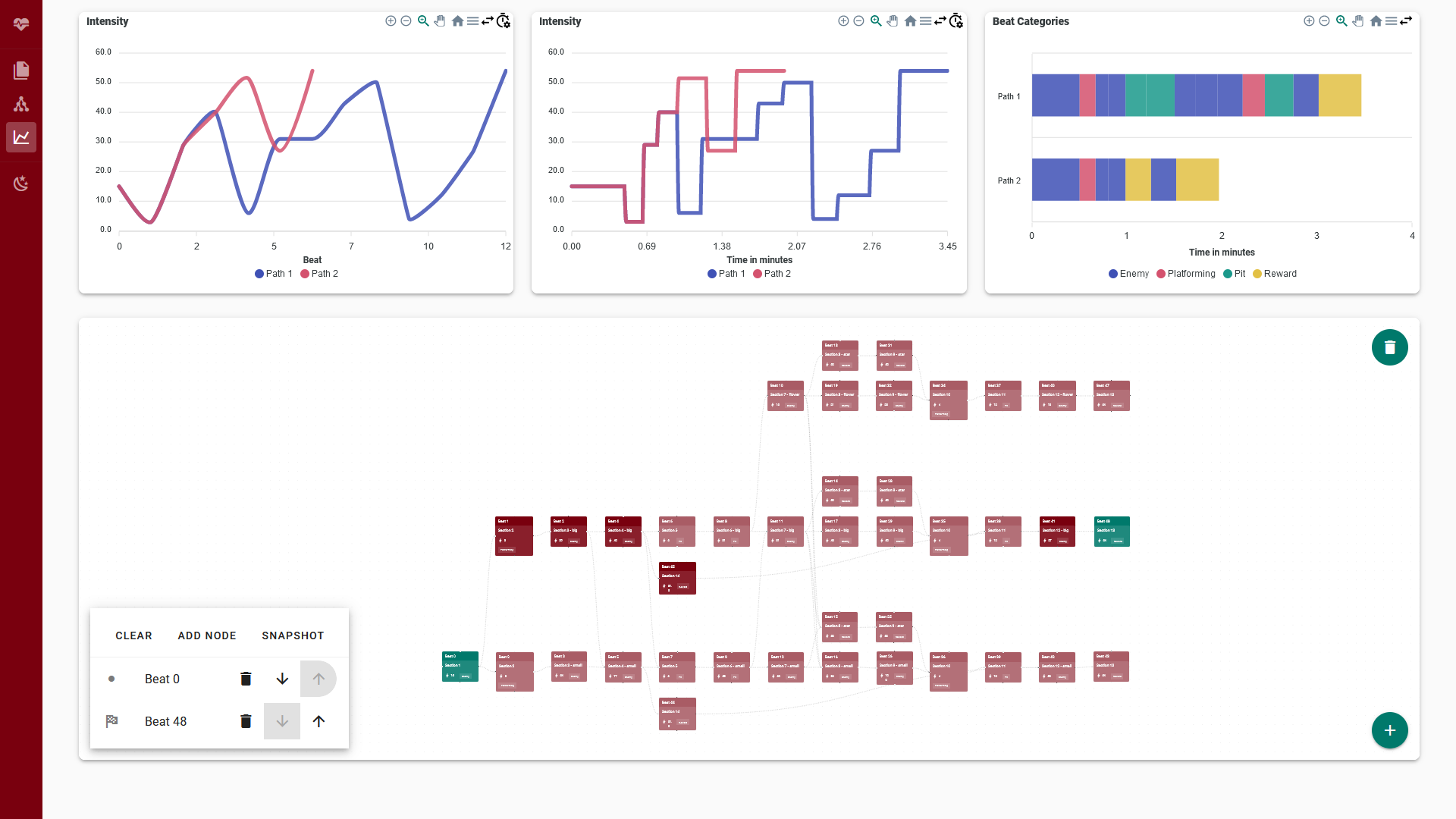}
    
    \includegraphics[trim=521 517 482 10, clip, width=0.33\textwidth]{sections/results/path_flower_neutral.png}
    \includegraphics[trim=970 517 33 10, clip, width=0.33\textwidth]{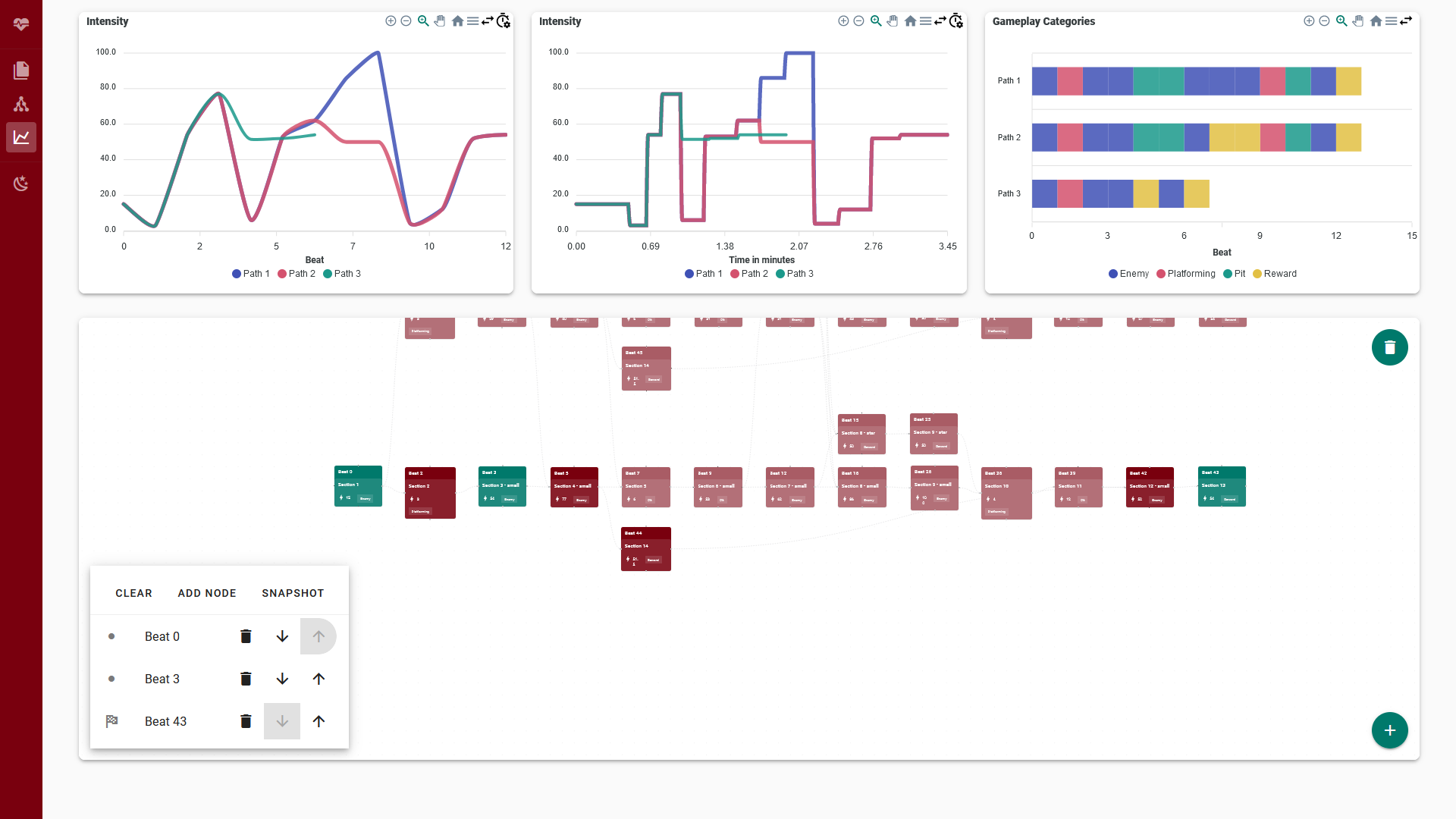}
    \includegraphics[trim=970 517 33 10, clip, width=0.33\textwidth]{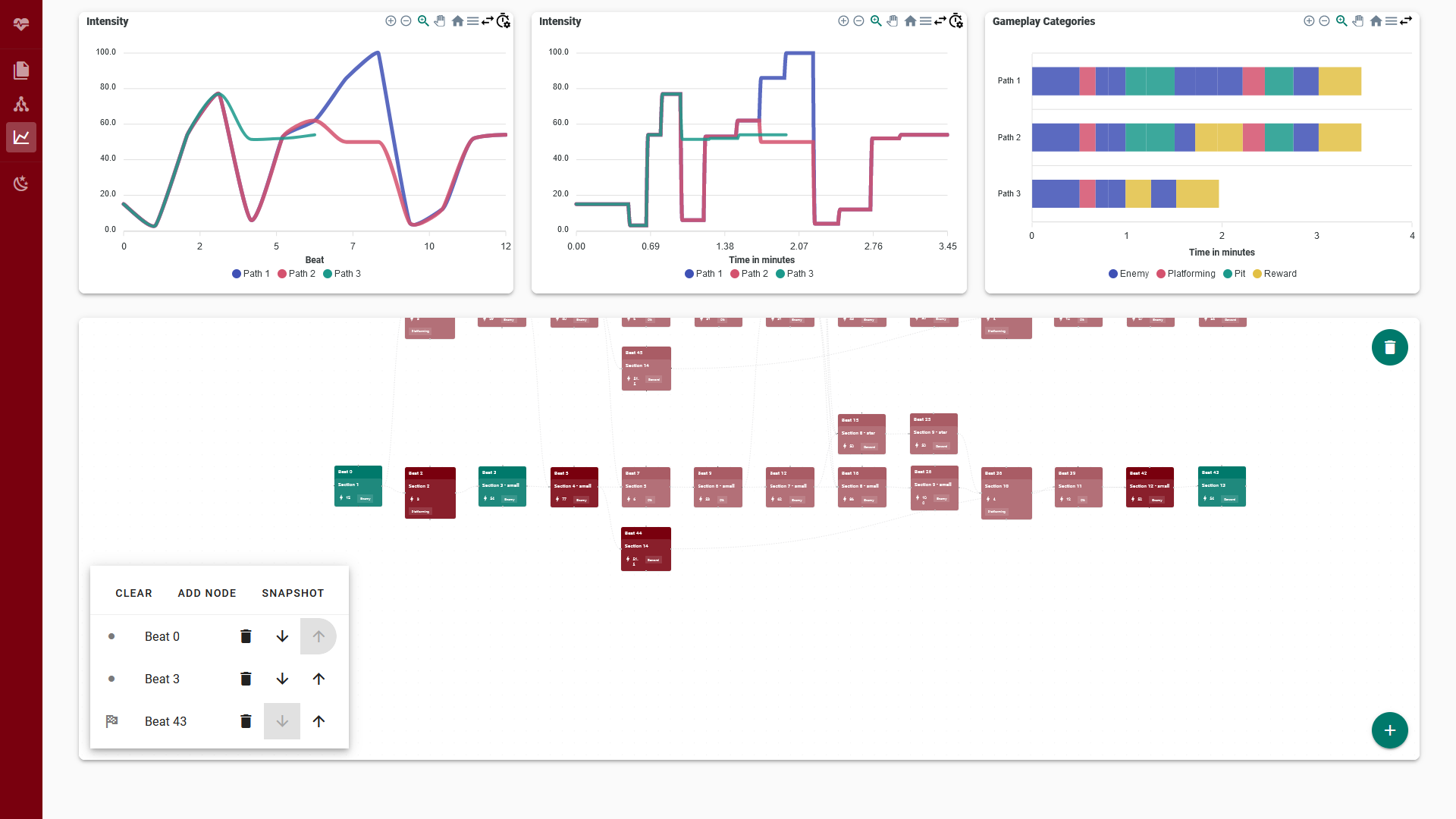}
    \caption{A showcase of PaceMaker's pacing diagrams of the demonstration with various settings. Top-left and bottom-left show the intensity diagram of the small--neutral route (path 1), the big--neutral route (path 2), and the flower--neutral route (path 3) with intensity settings \textit{Computed} and different time settings. Top-middle shows the big--star route with intensity setting \textit{All}. Top-right shows a comparison between the big--neutral route (path 1) and the big-hidden route (path 2) with intensity settings \textit{Computed}. Bottom-middle and bottom-right show a neutral route (path 1), a star route (path 2), and a hidden route (path 3) on the category chart with different time settings.}
    \label{fig:pacing_diagrams}
\end{figure*}

After the experience chart is created, the user can select paths to visualize the experience through pacing diagrams.
For this part of the demonstration, we call a path from the first to the last section a \textit{route}, and define relevant routes as follows:
\begin{itemize}
    \item[] \textbf{Neutral Route}: A route without any beat representing Section 8 and 9 in the star state or Section 14
    \item[] \textbf{Star Route}: A route with beats representing Section 8 and 9 in the star state
    \item[] \textbf{Hidden Route}: A route with a beat representing Section 14
    \item[] \textbf{Small Route}: A route without any beat representing the big or the flower player state
    \item[] \textbf{Big Route}: A route without any beat representing the small player state after the first power--up, and without any beat representing the flower player state
    \item[] \textbf{Flower Route}: A route without any beat representing the small player state after the first power--up, and without any beat representing the big player state after the second power--up
\end{itemize}
Everything else is a mixed route.
Since covering all possible routes would be going beyond the scope, we focus on the following selection to demonstrate intended use cases: the \textit{small--neutral route}, the \textit{big--neutral route}, the \textit{flower--neutral route}, the \textit{big--star route}, and the \textit{big--hidden route} (\autoref{fig:pacing_diagrams}).

\subsection{Expert Interviews}
To get initial insight into how how PaceMaker might fare in a real-world environment, interviews with two participants (P1, P2) were performed.
The interviews were conducted and recorded through an online video conferencing platform.
Each interview took about 50 minutes and was held in German.
In the beginning, we explained that the goal is to understand the potential of the toolkit in a working environment and that honest and direct answers are most valuable.
Afterward, the consent form was signed, and demographics were collected through a questionnaire.

\subsubsection{Demographics}
Both participants are male, live in Germany, have a college or university degree, and actively work in the game industry.
Further answers to the demographics form can be found in \autoref{table:demographics}.
During the interview process, we additionally asked about the team sizes:
\begin{itemize}
    \item[\textbf{P1:}] $\sim4$
    \item[\textbf{P2:}] $\sim50$, design department: $\sim10$
\end{itemize}

\subsubsection{Interview Process}
The first part of the interview consisted of asking the participants about pacing and workflows they might use to design pacing.
This was done before showing anything about PaceMaker to get unbiased answers.

After the introductory questions, PaceMaker was demonstrated by the interviewer through screen share.
A demonstration was chosen to avoid the participants being distracted by usability issues and instead focus on functionality and usefulness.
During the showcase, the complete workflow was shown by a minimal example task.
The task was to design the pacing of a level from an action-adventure game with five sections.
The level starts with a cinematic, followed by a platforming section.
Afterwards, a dialogue section gives the player two choices.
One choice leads to a puzzle section, the other to an action section.
Both finally lead to the same boss fight.
After creating the corresponding experience specifications and the experience chart, the functionality of the pacing diagrams was explained, and their initial impression was asked.
Subsequently, we showed the participants the prepared \textit{World 1-1} chart and diagrams for a bigger example and asked if their impression had changed.
Finally, we prepared some questions for after the initial impression of the participants, which focused on current features of PaceMaker.

\begin{table}
    \caption{Answers of each participant to the demographic form questions}
    \label{table:demographics}
    \centering
    \begin{tabular}{ | m{1em} | m{1.5em}| m{1.25cm} | m{5cm} | }
        \hline
         & Age & Job Title & Please describe in a few sentences what your typical working day looks like and what your tasks and responsibilities within the team are. \\ [0.5ex] 
        \hline

        \hline
        P1 & 26 & Game Director & Game Direction, communicating the project vision to team members.
        Project Planning, Planning resources, budget and the Asset Pipelines.
        Programming, Writing the code that the game consists of.
        Level Design, Building Levels and integrating Assets of Artists.
        \\
        \hline
        P2 & 27 & Quest and Content Designer & I plan and discuss concepts for quests and narrative gameplay moments in our current project. After that I start implementing them in the game engine and iterate on them depending on others' feedback.
        \\
        
        \hline
    \end{tabular}
\end{table}

\subsubsection{Summary}
Both participants' understanding of pacing fit our definition. 
While P1 did not seem too familiar with the concept, P2 is using pacing and pacing diagrams in his everyday work. 
He mentioned that he used to use them more often, but that changing while iterating is difficult and therefore does not fit well with the current workflow.
The first impression of both participants was that the amount of data the user has to input is too much or takes too long to be effective, indicating a weakness in the current workflow of the tool. 
P1 explained that he would rather use more coarse beats to evaluate the game on a higher level, which would also counteract the mentioned concern.
P2 expressed that he had been wondering throughout the demonstration, how the tool could be integrated into his workflow.
None of the current features stood out as particularly better or worse than the rest. 

\subsubsection{Conclusion}
To conclude the interviews, we asked the participants a final summarizing question and gave them the opportunity to voice anything else that they wanted to add.
The following is a paraphrased English translation of the original answers:

\textit{Imagine the final version of the tool with all the relevant features discussed today. Can you imagine yourself working with it at work?}
\begin{itemize}
    \item[\textbf{P1}] I think if I used the tool, then it would be for planning. In general, how the game should be structured, but then on a very high level. Like I said earlier, the levels back to back, and not the levels themselves. Like how should the entire game progress from start to finish.
        And I would include things as the title screen there. That is super important, the player is thrown into the game and has to be hooked immediately. So the first time a game is played, the player might end up in a special screen with little options and that node leads to a tutorial, while in the second playthrough you start in the main menu. I think those are the paths that you should consider and plot. That's what we considered for our last game, where we had to make sure it is not too intense in the beginning because it's the first thing the player sees. 
        Also, the flow within a world, e.g., you have level, level, level, followed by a cutscene, where the player has nothing to do for a minute, and then again level, level, boss fight. Something like that. And then you might notice that there should be a break before the boss fight because there are too many levels in between.
        That's how I would use it, but not within the levels.
    \item[\textbf{P2}] Actually yes, for sure. If everything goes by plan, and I have a proper planning phase, followed by an implementation phase, then this tool certainly provides additional value. I would use it for every quest and every level because it also takes work off my hands.
        I always try to do that myself, in Miro or Excel, for example. I take some values and input them. Then I have these separate beats, just like you do, and I assign some value. Then I have a bar chart and I connect the peaks of the bars, which creates a line. That's a huge effort for every quest and is also not very flexible.
        If I had the tool and set it up properly once, think about the parameters that are important to me, and it shows me the graphs and I can switch the settings and also connect it to testing and playing, that would make a lot of sense.
        The question is if it fits into the production style of the studio because things are significantly more chaotic and things are just there suddenly, and I am forced to work with the things I have. It's not like in the textbook, like we have these mechanics and this execution and this balancing, but instead, we have this level, make a quest for it. And I am still learning. 
        But even in this environment, it has its place.
\end{itemize}
\textit{Is there anything you want to add or do you have any questions?}
\begin{itemize}
    \item[\textbf{P1}] Not really.
    \item[\textbf{P2}] Actually, there is one more thing that I want to add. I mentioned that our production is a bit chaotic, but I am just starting to wonder if it is exactly because we are not using tools like this. Because we don't really have anything that goes into that direction, to plan quests or in an open--world game to plan the play experience or the golden path on a global scale.
    Everything is just done kind of intuitively. And that definitely creates moments where the intent of the designer is lost, and you start thinking some things could be better. Integrating those kinds of tools into the workflow could absolutely contribute to the fact that people are kind of forced to think about their decisions. So it would also be a benefit for companies. Not necessarily because it is the perfect tool, but rather because the usage of such a tool creates more structure and planning.
\end{itemize}
\textit{So you mean any tool is a good start?}
\begin{itemize}
    \item[\textbf{P2}] Yes. The more it takes work off your hands, and the more versatile it is, the better. Otherwise, you would just jump from one tool to the next for different things, and you don't want that either.
\end{itemize}

%% file: sections/discussion.tex
\section{Discussion}\label{sec:discussion}
The results indicate that this research successfully achieves its predefined objectives.
The demonstration provides an investigation of the toolkit's ceiling, and the initial interviews revealed existing interest in such a tool.
However, the interviews also identified issues of the current prototype, showing there is still room for improvement. 

As shown in the walkthrough, PaceMaker can handle various use cases and supports a minimal set of concepts that are essential for pacing considerations.
Furthermore, the demonstration showcases the toolkit's ability to handle complexity and potential solution space. 
Experience charts can model the non-linear game space, while pacing diagrams can visualize and compare linear sequences of beats to others. 
This enables users to dynamically modify and communicate requirements.
Our small qualitative study, consisting of interviews with game designers, supports these claims.
By analyzing their responses, we can better understand the strengths, weaknesses, and potential usefulness of the toolkit. 
P2's answers highlight the problem statement and suggest the need for tools like PaceMaker to enhance iterative workflows in game development and design.

However, the study's small sample size precludes any definitive conclusions.
Furthermore, during the interviews, concerns about usability arose despite our use of a demonstration.
Usability and a low entry barrier are critical to the success of any tool, and these factors were not considered in our first prototype.
Additionally, the limitations of PaceMaker in handling complexity may not be immediately evident.
According to Harel~\cite{harel1987statecharts}, nesting and concurrency become necessary with increasing amounts of complexity.
Therefore, this should also be necessary for larger examples in our experience chart.
Finally, it is unclear how and if the tool can be integrated into the actual game production workflow. 

Overall, the results suggest that PaceMaker is on the right track to being a productive toolkit for game designers and game development companies.
It has a high ceiling, is flexible with regard to the level of detail, and can showcase many different valuable diagrams and data sets.
However, expected concerns about the path of least resistance and general usability arose during the interview process, which hinder the toolkit from entering a production environment.
To simplify the process, an interface can be implemented between PaceMaker and game engines. 
This will allow data to be fed directly into PaceMaker from within the game.
Furthermore, we suggest to conduct a field study and collaborate more closely with the target group to gain insight into the intricacies of their workflows.
Nevertheless, these problems were not the focus of our research, and are not indicative of PaceMaker's potential and expressiveness.

Therefore, the current state of PaceMaker represents a significant advancement towards a useful solution to the problem statement.